\documentclass[sigconf]{acmart}
\AtBeginDocument{%
  }

\setcopyright{acmlicensed}
\copyrightyear{2018}
\acmYear{2018}
\acmDOI{XXXXXXX.XXXXXXX}
\acmConference[Conference acronym 'XX]{Make sure to enter the correct
  conference title from your rights confirmation email}{June 03--05,
  2018}{Woodstock, NY}
\acmISBN{978-1-4503-XXXX-X/2018/06}

\usepackage{siunitx}

\begin{document}

\title{Shifting in-DRAM}

\orcid{1234-5678-9012}
\author{William C. Tegge}
\email{wtegge@syr.edu}
\affiliation{%
  \institution{Syracuse University}
  \city{Syracuse}
  \state{New York}
  \country{USA}
}


\author{Alex K. Jones}
\email{akj@syr.edu}
\affiliation{%
  \institution{Syracuse University}
  \city{Syracuse}
  \state{New York}
  \country{USA}
}

\renewcommand{\shortauthors}{Trovato et al.}

\begin{abstract}
Processing-in-Memory (PIM) architectures enable computation directly within DRAM and help combat the memory wall problem. Bit-shifting is a fundamental operation that enables PIM applications such as shift-and-add multiplication, adders using carry propagation, and Galois field arithmetic used in cryptography algorithms like AES and Reed-Solomon error correction codes. Existing approaches to in-DRAM shifting require adding dedicated shifter circuits beneath the sense amplifiers to enable horizontal data movement across adjacent bitlines or vertical data layouts which store operand bits along a bitline to implement shifts as row-copy operations. In this paper, we propose a novel DRAM subarray design that enables in-DRAM bit-shifting for open-bitline architectures. In this new design, we built upon prior work that introduced a new type of cell used for row migration in asymmetric subarrays, called a “migration cell”. We repurpose and extend the functionality by adding a row of migration cells at the top and bottom of each subarray which enables bidirectional bit-shifting within any given row. This new design maintains compatibility with standard DRAM operations. Unlike previous approaches to shifting, our design operates on horizontally-stored data, eliminating the need and overhead of data transposition, and our design leverages the existing cell structures, eliminating the need for additional complex logic and circuitry. We present an evaluation of our design that includes timing and energy analysis using NVMain, circuit-level validation of the in-DRAM shift operation using LTSPICE, and a VLSI layout implementation in Cadence Virtuoso. 

\end{abstract}




\maketitle

\section{Introduction}
\begin{sloppypar}
The memory wall problem refers to the growing disparity between memory bandwidth and processing speeds and has resulted in increasing interest in Processing-in-Memory (PIM) architectures \cite{wulf1995hitting}. PIM eliminates costly data transfers between the processor and main memory and reduces latency and energy consumption by performing computation directly in memory. Researchers have shown that bulk bitwise operations like AND, OR, and NOT can be efficiently performed in DRAM subarrays. This is done by exploiting charge sharing on the bitlines and changing the sense amplifier behavior \cite{seshadri2017ambit}. Bit-shifting stands out among the fundamental operations required for PIM workloads. Shifting enables a wide range of arithmetic and logical computations that could not be done without moving data off-chip for processing. Addition with carry propagation, when implemented in a bit-serial fashion, benefits from shifting. A common multiplication algorithm, shift-and-add multiplication, relies on repeated shift operations to align partial products before the accumulation step \cite{de2024count2multiply}. Galois field arithmetic depends on shifting for the polynomial multiplication and reduction. Galois field operations form the core of widely deployed algorithms in cryptography such as Advanced Encryption Standard (AES) \cite{daemen2002design} and in Reed-Solomon error correction codes used in communication protocols \cite{reed1960polynomial, wicker1999reed}. Having an efficient in-DRAM shifting operation would open up opportunities for acceleration in various applications such as cryptography, machine learning, and data processing.

Implementing in-DRAM bit-shifting presents challenges that do not arise with the regular bitwise logic operations. In a standard DRAM subarray, each bitline is electrically isolated from its neighbors. Cells within a column can share charge between each other and the sense amplifier. This charge sharing along a bitline enables row copy \cite{seshadri2013rowclone} and bulk AND/OR operations \cite{seshadri2017ambit}. Currently, there is no native mechanism that allows for horizontal data transfer between adjacent bitlines. DRAM columns are hardwired to their respective sense amplifier circuits. This constraint makes bit-shifting more difficult to implement than operations that act on each column independently.

Prior works have attempted to address this challenge. SIMDRAM employs a layout in which all bits of an operand are stored vertically along a single bitline instead of horizontally across a row \cite{hajinazar2021simdram}. In this layout, bit-shifting is reduced to a row-copy operation, and this row-copy operation can be done using an existing mechanism like RowClone \cite{seshadri2013rowclone}. However, SIMDRAM introduces significant overhead. Data must be transposed from the conventional horizontal layout to the vertical layout before computation and then transposed back afterwards. This method incurs energy and latency penalties, requires dedicated transposition hardware, and complicates system integration due to the vertical layout being incompatible with standard memory access patterns. DRISA enables horizontal data movement between adjacent bitlines by adding dedicated shifter circuits beneath the sense amplifiers \cite{li2017drisa}. While this does not require data transposition, it does require additional transistors and wiring in an area of the chip that is already constrained. The additional circuitry increases complexity and area overhead.

In this paper, we propose a novel approach to in-DRAM bit-shifting for open-bitline architectures using migration cells. Prior work introduced migration cells originally for row migration between near and far segments of a dynamic asymmetric subarray \cite{lu2015improving}. We repurpose and extend this idea by placing a row of migration cells both at the top and bottom of the subarray which enables bidirectional bit-shifting within a given row. This approach operates on horizontally-stored data eliminating the need and associated overhead of data transposition.  Using migration cells takes advantage of the existing cell structure, minimizes additional circuity, and maintains compatibility with standard DRAM operations. 
\end{sloppypar}

We make the following contributions:
\begin{itemize}
\item We propose a novel DRAM subarray design that enables in-DRAM bi-directional bit-shifting for open-bitline architectures using migration cells at the top and bottom of the subarray. 
\item We analyze timing and energy of the proposed shift operation using NVMain.
\item We validate the circuit-level operation and reliability of the shift using LTSPICE simulations.
\item We demonstrate the physical feasibility of our proposed design through a VLSI layout implemented in Cadence Virtuoso. 
\end{itemize}

\section{Background}
\label{sec:Background}

\begin{sloppypar}
First, we briefly review DRAM organization and architecture, subarray structure, and PIM operations (AND, OR, NOT, RowClone, and multi-row Activation) of COTS DRAM chip. 
\end{sloppypar}

\subsection{DRAM Organization}

Modern DRAM systems are organized hierarchically, from the memory controller to the individual memory cells. At the system level, a memory controller communicates with DRAM through one or more memory channels, and each channel connects to one or more ranks. Each rank consists of multiple DRAM chips operating together. Inside each DRAM chip, there are centralized I/O pads that receive and transmit command and data signals over the external memory bus. The commands are decoded and forwarded to the appropriate bank based on the address. Each chip is divided into multiple banks, and these banks can operate in parallel to increase memory-level parallelism. Within each bank, memory is organized as a 2-dimensional array of DRAM cells. The bank includes peripheral logic that selects a wordline (row logic) and selects specific columns from the activated row (column logic). Instead of being built as one large array, each bank is split into multiple subarrays to limit wire length and improve performance. Each DRAM cell consists of a storage capacitor that stores data as electrical charge and an access transistor that is controlled by a wordline. The access transistor connects the capacitor to the bitline. The intersection of a wordline (row) and a bitline (column) forms a DRAM cell, which stores a single bit of information.

\subsection{Subarray Structure}
A DRAM bank is divided into many subarrays. Each subarray contains a 2-dimensional array of DRAM cells, sense amplifiers (that are also referred to as row buffers), and local row decoders. Modern DRAM uses an architecture called open-bitline. Open-bitline architecture reduces cell area (from 8F\(^2\) to 6F\(^2\)) and improves density. In open-bitline, each bitline is connected to a sense amplifier shared between two neighboring subarrays. When a row is activated, the row is split between the top and bottom row buffers to be read out. Access transistors turn on when the wordline connected to it is activated. After the wordline is activated, the charge in the storage capacitor is shared with the bitline. The bitlines have large capacitances, so the resulting voltage has a small deviation. The sense amplifier detects and amplifies the voltage difference to a full logic level of 0 or VDD, and the full charge is stored back into the cell. 
\subsection{PIM Operations}
A complete DRAM access has three stages: ACTIVATE (ACT), READ/WRITE (RD/WR), and PRECHARGE (PRE). During an ACT, the wordline is raised allowing charge from all cells in the row to flow onto the bitlines. After the charge sharing is complete, the sense amplifiers detect and amplify the voltage difference, The amplified value is restored back into the cells. For a RD, column logic selects specific columns from the activated row, and the data is transferred between the row buffers and the external memory channel. For WR, new data is driven into the row buffer and then restored into the cells. PRE prepares the bank for the next ACT command by driving bitlines to approximately \(1/2\)VDD.

Processing-in-memory (PIM) techniques exploit charge sharing and sense amplifiers inside DRAM subarrays to operate on data. Ambit \cite{seshadri2017ambit} has shown that PIM operations are possible through RowClone and Multi-Row Activations (MRA) and that these PIM operations are functionally complete. RowClone \cite{seshadri2013rowclone} performs data copy between rows within a subarray using an ACT-ACT-PRE (AAP) sequence. Both rows are activated in sequence, and since those cells share the same sense amplifiers, the charge stored in the row buffer can be written directly into the destination row. This enables fast, high-bandwidth in-DRAM copying. Multi-Row Activation (MRA) is an ACT-PRE command sequence. The first ACT can be a Dual-Row activate (DRA) or a Triple-Row activate (TRA). TRA can be used to implement bitwise logic operations like MAJ, AND, and OR. MRA operations are destructive which means at the end of the command sequence, all activated rows are overwritten with the final result. Ambit implements bitwise NOT using dual-contact cells (DCCs). DCCs consist of two access transistors controlled by different wordlines. The upper transistor connects the storage node to the bitline for a standard read/write operation, and the lower transistor provides a path that supports logical NOT. 

In open-bitline architecture, bitwise NOT operations can be achieved naturally by moving charge across a subarray from one subarray to another. Moving a charge across the shared sense amplifier results in the logical inversion of that charge being written to the destination row in the adjacent subarray.

\section{Proposed Design}
\label{sec:Proposed Design}

In this section we review the migration cell design, present our subarray design with the migration cells and shifting capabilities, and we discuss the procedure for a 1-bit shift operation.  

\subsection{Migration Cell Circuit Design}

The migration cells were originally purposed to do row migration between subarrays \cite{lu2015improving}. The special cell has two access ports that share a single storage capacitor. This structure allows two unassociated bitlines to access the shared storage cell. It acts as temporary storage both for migration and shifting. This cell was designed in this way because using a simple transistor between the two adjacent bitlines greatly increases overhead, harshly impacts layer density rules, and degrades yield \cite{lee2013tiered}. In section 6, we present and discuss the corresponding VLSI layout of a migration cell used in our design.

\subsection{Subarray Design}

Our new subarray design uses two migration cell rows in each subarray. We place one at the top row and one at the bottom row. Each migration row is connected to two wordlines to allow for the charge-sharing and bit-shifting mechanisms to happen correctly. Figure 1 shows this new layout. 

\begin{figure}
    \centering
    \includegraphics[width=1\linewidth]{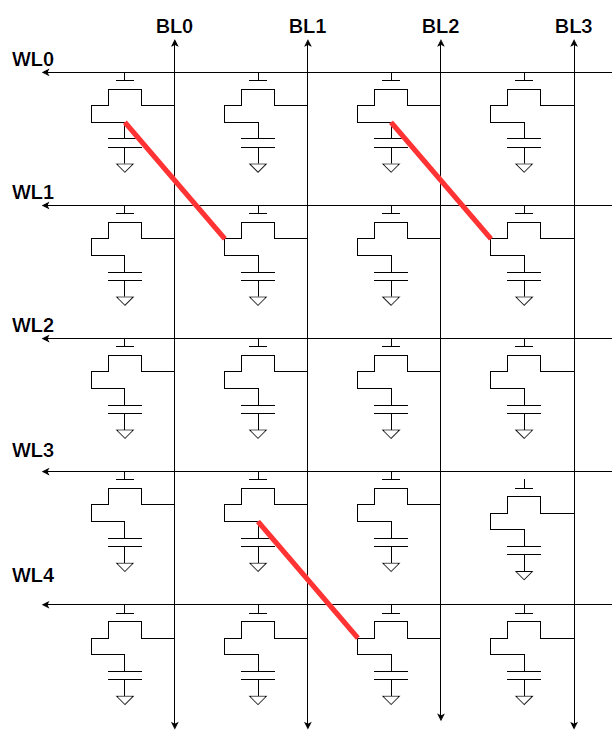}
    \caption{Circuit Schematic of Standard DRAM Cells and Migration Cells}
    \label{fig:placeholder}
\end{figure}

Figure 2 shows why one row of migration cells is not enough and the 2nd row of migration cells is necessary for bit shifting a full row. If we used the original DAS-DRAM design \cite{lu2015improving}, bits would only be able to shift by 1 position. The bits in odd-numbered columns would only be able to shift left by 1 position, and the bits in even-numbered columns would only be able to shift right by 1 position. When this shift occurs, the data in the dst position gets overwritten. 

\begin{figure}
    \centering
    \includegraphics[width=1\linewidth]{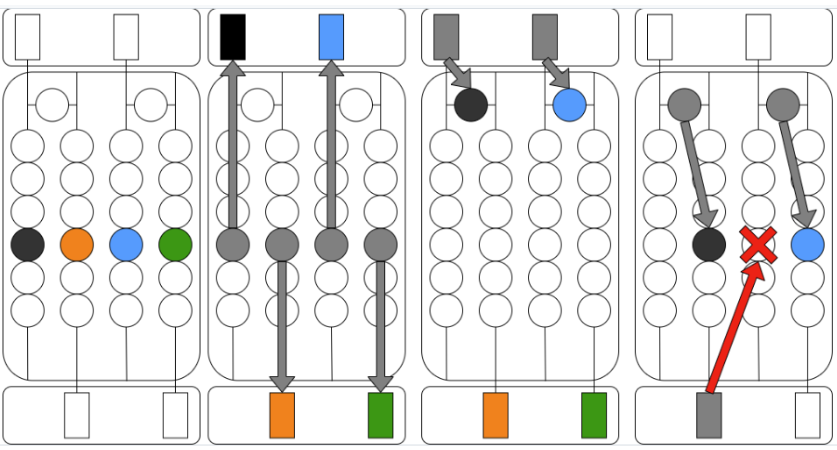}
    \caption{Shift With Only One Row of Migration Cells}
    \label{fig:placeholder}
\end{figure}

\begin{figure}
    \centering
    \includegraphics[width=1\linewidth]{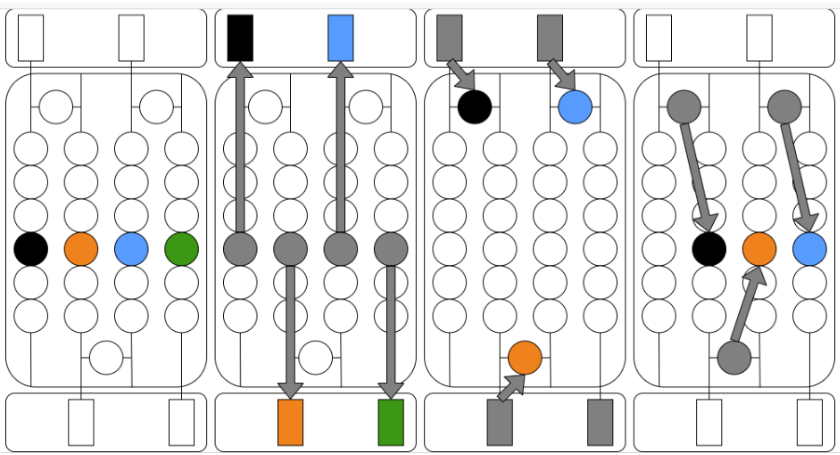}
    \caption{Shift With Two Rows of Migration Cells}
    \label{fig:placeholder}
\end{figure}

\subsection{Shift Procedure}
Figure 3 shows the procedure for a full row 1 bit right shift. The shift procedure takes 4 AAP commands (essentially 4 RowClones) in sequence. For a right shift, the first row clone happens between the src row to the top migration row followed by another row clone from src to the bottom migration row. At this point, the src row is split between the top and bottom migration rows. The top migration row holds all the bits from the even numbered columns of the src row, and the bottom migration row holds all the bits from the odd numbered columns. The top migration row is then copied to the dst row, and in the final step, the data is combined after the bottom migration row is copied to dst.  

For a right shift, the bits in the even columns (columns 0, 2, etc) of the row will go up to the top migration row, and the bits in odd columns will go down to the bottom migration row. However, when performing a left shift, the bits in the even columns go to the bottom migration row, and the bits in the odd columns go to the top migration row. This means that the sequence of row clones and wordlines that are activated during the process is different depending on which way you are shifting.

\section{Experimental Setup}
\label{sec:Experimental Setup}

We present a multi-faceted methodology that consists of architectural simulation, circuit-level validation, and a physical layout implementation to evaluate the energy consumption, performance, area overhead, and circuit-level reliability of our proposed migration-cell based shifting design. This section lists the tools, workloads, and configurations used in our evaluation. 

\subsection{Architectural Simulation}

We use a cycle-accurate memory simulator called NVMain \cite{poremba2012nvmain} to evaluate the timing and energy characteristics of our in-DRAM shift operation. NVMain models DRAM at the command level and provides detailed and accurate energy breakdowns for different DRAM operations such as activation, read, and write and PIM primitives such as SRA, DRA, and TRA.  

We configure NVMain to model a Micron DDR3-1333 4Gb DRAM chip with the following parameters:
\begin{itemize}
    \item Memory technology: DDR3-1333 (667 MHz data rate)
    \item Chip capacity: 4Gb
    \item Bank organization: 8 banks per rank, 2 ranks per channel, 2 channels
    \item Subarray structure: 512 rows per subarray, 8KB row buffer size
    \item Timing parameters: Standard DDR3-1333 timing (tRCD=13.5 ns, tRP = 13.5 ns, tRAS = 36 ns, tRC = 49.5 ns, tREFI = 7.8 \unit{\us})
\end{itemize}

This configuration is a typical commercially available DDR3 DRAM implementation and provides a realistic baseline for evaluating our design. 

We evaluate four workloads with varying numbers of shift operations in order to assess both baseline performance and scalability:
\begin{itemize}
    \item 1 shift: 1 shift operation (baseline measurement)
    \item 50 shifts: Small workload to observe refresh impact
    \item 100 shifts: Medium workload
    \item 512 shifts: Large workload to validate scalability
\end{itemize}

Each shift operation shifts all bits in a full 8KB row (65,536 bits) by one position. All shift operations process data stored in conventional horizontal layout, with each byte occupying consecutive bitlines within a row. The workloads execute sequentially within Bank 0, allowing us to measure single-bank performance characteristics.   

NVmain breaks energy consumption down into several categories: 
\begin{itemize}
    \item Active energy: Energy consumed during row activation (AAP commands)
    \item Burst energy: Energy for data transfer on/off chip
    \item Refresh energy: Background refresh operations
    \item Precharge energy: Energy for closing rows
    \item Standby energy: Background idle power
\end{itemize}

We focus on active energy and burst energy because these represent the dominant components for PIM operations. The reported energy value include all activity within Bank 0 Subarray 0. 

\subsection{Circuit-Level Validation}

We validate our shift mechanism using LTSPICE, an industry-standard SPICE circuit simulator. Our LTSPICE simulation verifies that the charge-sharing mechanism between migration cells and standard DRAM cells correctly implements the bit-shift operation at the transistor level. 

Our LTSPICE simulations models:
\begin{itemize}
    \item Standard 1T1C DRAM cells with realistic capacitance values and access transistor parameters based on the technology node
    \item Migration cells 
    \item Bitline parasitic capacitance and wordline signal delay
    \item Sense amplifier circuits for read operations
    \item Control signal timing for AAP command sequences
\end{itemize}

We perform circuit-level validation of four different technology nodes to ensure that our design remains viable across current and future DRAM process technologies. We use the following technology nodes:
\begin{itemize}
    \item 45nm: Using PTM (Predictive Technology Model) transistor parameters
    \item 22nm: Using PTM model transistor parameters
    \item 20nm: Estimated by scaling the device parameters and making proportional adjustments to transistor width, length, threshold voltage, and parasitic capacitances from the 22nm model. 
    \item 10nm: Estimated by scaling from both the 22nm and 20nm models
\end{itemize}

The PTM models \cite{websitekey} provide realistic transistor behavior including short-channel effects, velocity saturation, and process variations that are critical for accurate DRAM circuit simulation. Standardized PTM models were not directly available for the 10nm and 20nm nodes, so we estimated device parameters by applying scaling factors to threshold voltages, gate capacitances, and transistor dimensions based on the already established 22nm and 45nm models \cite{zhao2006new} \cite{cao2011predictive}. Our scaling methodology follows industry-standard practices for predictive device behavior at advanced nodes. Table 1 shows the values used for various parameters at each technology node.

\begin{table}
\centering
\footnotesize
\begin{tabular}{| l | l | l | l | l | l | l |}
\hline
\textbf{Parameter } & \textbf{600nm} & \textbf{180nm} & \textbf{45nm} & \textbf{22nm} & \textbf{20nm} & \textbf{10nm} \\
\hline
\textbf{Vdd} & 3.3V & 1.8V & 1.5V & 1.2V & 1.1V & 1.1V \\
\hline
\textbf{WL boost} & 5.0V & 3.3V & 3.0V & 2.5V & 2.4V & 2.2V \\
\hline
\textbf{Cell Cap} & 120fF & 50fF & 30fF & 25f & 25fF & 18fF \\
\hline
\textbf{Access L} & 0.6u & 0.18u & 0.045u & 0.022u & 0.020u & 0.012u \\
\hline
\textbf{Access W} & 1.2u & 0.36u & 0.18u & 0.044u & 0.040u & 0.025u \\
\hline
\textbf{SA NMOS W} & 140u & 42u & 10.5u & 7u & 6u & 4.5u \\
\hline
\textbf{BL R/cell} & 1000m & 400m & 200m & 120m & 110m & 100m \\
\hline
\textbf{BL C/cell} & 2.0f & 0.8f & 0.40f & 0.24f & 0.22f & 0.18f \\
\hline
\textbf{trise} & 5n & 2n & 0.7n & 0.5n & 0.4n & 0.3n \\
\hline

\end{tabular}
\caption{DRAM cell and circuit parameters across technology nodes used in LTSPICE simulations}
\end{table}

We simulate both right-shift and left-shift operations to validate the bidirectional capability of our design. For each shift direction and technology node, we simulate the four-step AAP sequence required for a single bit-shift operation. We use varied data patterns (all zeros, all ones, alternating, and random) to ensure operation correctness. 

Our simulations validate several critical properties:
\begin{itemize}
    \item Successful data transfer
    \item Correct shift operation: the bit value at position i correctly shifts and appears at position i$\pm$1 after the shift sequence completes
    \item Data preservation in surrounding cells: Cells not directly involved in the shift operation (rows above and below the active row) retain their original values
    \item Signal integrity: Voltage levels remain within acceptable margins, with sense amplifiers correctly distinguishing between logical '0' and '1' states 
    \item Proper charge transfer: Migration cells successfully capture data and transfer the charge to adjacent bitline during the shift
    \item Complete write-back: Shifted data is successfully written back to destination cells with sufficient charge for reliable retention until the next refresh cycle
\end{itemize}

We use transient analysis with a 1 ns time step to capture the dynamic behavior of charge sharing during row activations. 

\subsection{Physical Layout Implementation}
We implement our VLSI layout for the proposed migration cell-based shifting design using Cadence Virtuoso Layout Suite, a comprehensive platform for custom integrated circuit design and verification. Our layout was developed using a 22nm technology node. Our design follows the strict technology specific design rules in order to ensure it will be manufacturable and reliable. The design rule checking (DRC) constraints included minimum width and spacing requirements for all the metal and diffusion layers and via enclosure rules to prevent process-induced damage. The wordline and bitline pitch constraints \cite{auth201222nm} were followed which allowed for efficient array tiling and minimizes routing complexity. 

We followed a hierarchical design approach in which we designed and optimized the migration and normal DRAM cells first, and then we fit them into the subarray structure. This design methodology allowed for systematic optimization of the area efficiency. 

\section{Evaluation}

In this section, we present a comprehensive evaluation of our migration cell-based in-DRAM shifting design. We analyze the energy and performance characteristics using architectural simulation (Section 5.1), validate circuit-level operation reliability in our LTSPICE simulations (Section 5.2), and examine the area overhead and compare it with prior PIM architectures (Section 5.3). Our evaluation shows that the proposed design achieves efficient bit-shifting while maintaining compatibility with standard DRAM operations and minimizing area overhead. 

\subsection{Energy and Throughput Analysis}
In this section, we evaluate the energy and performance characteristics of our proposed in-DRAM shift operation using NVMain \cite{poremba2012nvmain}, a cycle-accurate memory simulator. Our evaluation uses the Micron DDR3-1333 4Gb configuration with 8KB row size. This represents a typical DDR3 DRAM implementation. We simulate workloads ranging from 1 shift operation to 512 shift operations to assess the baseline performance and scalability of our design.

\begin{table}
\centering
\footnotesize
\begin{tabular}{| l | l | l | l | l |}
\hline
  & \textbf{Single Shift} & \textbf{50 Shifts} & \textbf{100 Shifts} & \textbf{512 Shifts} \\
\hline
Total Energy & 31.321 nJ & 1592.52 nJ & 3223.6 nJ & 16554.6 nJ \\
\hline
Active Energy & 30.24 nJ & 1515.4 nJ & 3030.81 nJ & 15513.5 nJ \\
\hline
Burst Energy & 0 nJ & 0 nJ & 0 nJ & 0 nJ \\
\hline
Refresh Energy & 0 nJ & 77.1171 nJ & 192.793 nJ & 1041.08 nJ \\
\hline
Energy Per Shift & 31.321 nJ & 31.85 nJ & 32.236 nJ & 32.333 nJ \\
\hline

\end{tabular}
\caption{Energy consumption Breakdown For Shift Operations Across Different Workload Sizes in Bank 0 Subarray 0}
\end{table}

\subsubsection{Energy Consumption}
The energy consumption of our shift operation is dominated by active energy, which corresponds to the energy required for row activations during the AAP (Activate-Activate-Precharge) command sequences. Each shift operation performs 4 AAP commands which activates one row and then another row to enable charge-sharing. This active energy (30.24 nJ for a single shift, scaling proportionally with workload size) accounts for 96-97\% of the total energy in Bank 0 Subarray 0. For all workloads (1, 50, 100, and 512 shifts), the burst energy is zero. This shows that our PIM approach performs computation entirely within the DRAM array without moving data off-chip. The energy per shift remains consistent at approximately 31-32 nJ across all workloads, showing excellent energy scalability. After normalizing by the amount of data processed (8KB per shift), our design achieves an energy efficiency of approximately 4 nJ/KB. This measurement remains consistent across workload sizes, only varying from 3.915 nJ/KB to 4.041 nJ/KB (a variation of only 3.2\%). The values for each energy category can be seen in Table 2. 

\begin{table}
\centering
\footnotesize
\begin{tabular}{| l  |l  |l  |l  |l |}
\hline
\textbf{Metric} & \textbf{Single Shift} & \textbf{50 Shifts} & \textbf{100 Shifts} & \textbf{512 Shifts} \\
\hline
Total Time & 208.7 ns & 10.291 µs & 20.733 µs & 106.272 µs \\
\hline
Latency per Shift & 208.7 ns & 205.8 ns & 207.3 ns & 207.6 ns \\
\hline
Throughput (MOps/s) & - & 4.86 & 4.82 & 4.82 \\
\hline
Throughput (GOps/s) & - & 0.00486 & 0.00482 & 0.00482 \\
\hline

\end{tabular}
\caption{Performance Characteristics of in-DRAM Shift Operations Evaluated in Bank 0}
\end{table}

\subsubsection{Performance Characteristics}
Table 3 shows the performance metrics for Bank 0 across all evaluated workloads. The latency per shift operation stays consistent, ranging from 205.8 ns to 208.7 ns. This near-constant latency demonstrates that our shift operation scales linearly with minimal overhead. The single shift baseline of 208.7 ns establishes the cost of executing four AAP commands required for one shift operation, and this latency is consistent with the latency estimated for a single AAP command ($\sim$ 49ns) \cite{seshadri2017ambit}. 

The throughput of approximately 4.82 million operations per second (0.00482 GOps/s) is consistent across the multi-shift workloads which confirms the predictable and stable performance of our shift operation.  This consistency shows that the shift operation does not suffer from performance degradation as the workload scales, a characteristic that is crucial for applications that require large data processing. 

\subsubsection{Refresh Overhead}
It is important to consider the impact of DRAM refresh operations in an energy analysis. For 1 shift operation (208.7 ns), no refresh occurs, resulting in an energy per shift at 31.321 nJ. One observation we make is that as simulation duration increases, periodic refresh operations begin to contribute more to the total energy consumption: 4.8\% for 50 shifts, 6.0\% for 100 shifts, and 6.3\% for 512 shifts. Refresh energy increases the absolute energy efficiency of the shift operation itself. Because this refresh overhead is inherent to DRAM technology, it does not fundamentally alter the energy efficiency of the shift operation itself and would be present in any DRAM-based PIM approach operating over similar timing. 

\subsubsection{Bank-Level Parallelism}
Another advantage of our design is the potential for bank-level parallelism. Our DRAM configuration contains 8 banks per rank, and each bank can operate independently. The shift operations are confined to a single subarray and do not require inter-bank communication, which means multiple shift operations can be executed in parallel across different banks. Our current evaluation only shows single-bank performance. By exploiting bank-level parallelism, the aggregate throughput can scale linearly with the number of banks. With 8 banks operating in parallel, the theoretical peak throughput for a single rank increases from 4.82 MOps/s to 38.56 MOps/s (0.03856 GOps/s). For a dual-rank configuration with 2 channels (as in our simulation), when all 32 banks (8 banks × 2 ranks × 2 channels) are used, the entire system could theoretically achieve 154.24 MOps/s (0.154 GOps/s). 

Bank-level parallelism is advantageous for applications that need to perform shift operations on multiple independent data elements. For example:
\begin{itemize}
    \item Vector operations: SIMD-style shifting on multiple vector elements stored in different banks
    \item Cryptographic operations: Multiple AES encryption rounds can be processed concurrently
    \item Parallel shift-and-add multiplication: Multiplication operations on separate operands can be distributed across banks
\end{itemize}

The energy efficiency per bank would remain constant regardless of how many banks are active in parallel since each bank's energy consumption is independent. Parallelization across 8 banks increases throughput 8× while maintaining the same energy per operation. This results in the same overall power consumption per unit of work.

\subsubsection{Comparison with Standard Data Movement}
Our in-DRAM shift operation eliminates the need for data movement between memory and the CPU. The normal approach would be to read the 8KB row from DRAM, perform the shift in the CPU, and write back the result into memory. Assuming DDR3 energy costs of $\sim$10-15 nJ per 64-byte transfer, moving 8KB results in 128 transfers which would consume 1,280-1,920 nJ for the read alone, plus a similar amount to write it all back. Our in-DRAM approach achieves the same result using only 31-32 nJ. This shows a 40-60\% reduction in energy when compared to the conventional data movement. This energy reduction is maintained when bank-level parallelism is exploited. 

\subsubsection{Comparision with Prior PIM Shifting Approaches}

In this section, we compare our migration cell-based design with three notable prior works in PIM: SIMDRAM \cite{hajinazar2021simdram}, DRISA \cite{li2017drisa}, and Ambit \cite{seshadri2017ambit}. 

Comparison with Ambit: Our design builds upon Ambit's charge-sharing mechanism for bulk bitwise operations. Ambit showed that triple-row activation (TRA) operations enable AND/OR operations with energy costs of $\sim$3-5 nJ/KB \cite{seshadri2017ambit}. Our shift operation achieves a similar energy efficiency of 4 nJ/KB. This is expected as both designs use the same underlying charge-sharing mechanism with multi-row activations. Ambit performs column-wise operations while our design extends to allow horizontal data movement through the migration cells. 

As mentioned previously, the latency of $\sim$207 ns per shift for an 8KB row is comparable to the baseline AAP mechanism discussed in the Ambit paper. Our design maintains Ambit's advantages: it operates on standard DRAM without the need of specialized compute units, is compatible with other ambit-based operations, and works with commercially available DRAM chips with minor modifications. 

Comparison with SIMDRAM: While SIMDRAM can perform a shift in only a single row-copy operation ($\sim$50-100 ns using mechanisms like RowClone \cite{seshadri2013rowclone}), it incurs overhead from the mandatory data transposition. Each operand must be transposed from horizontal to vertical layout before computation and then transposed back afterwards. This transposition involves thousands of individual read and write operations for an 8KB row. SIMDRAM reports transposition latencies ranging from several microseconds to tens of microseconds depending on the data size. The energy costs can exceed 1,000-10,000 nJ for large operands. This transposition energy alone is 100-300× higher than our design's entire shift operation cost of 31-32 nJ. Our design eliminates the transposition overhead completely by operating directly on conventional horizontally-stored data which keeps it compatible with standard memory operations.  

Comparison with DRISA: DRISA enables horizontal shifting by adding dedicated shifter circuits beneath the sense amplifiers \cite{li2017drisa}. These shifter circuits include transistors, multiplexers, and wiring to physically move data between adjacent bitlines. DRISA reports energy per shift operation to be $\sim$5-20 nJ and shift latencies of $\sim$20-40 ns per bit position shifted. DRISA adds substantial area overhead in order to achieve this improved speed and energy. The shifter circuits must be replicated for every bitline. DRISA reports area overhead of $\sim$5-10\% depending on configuration. In contrast, our design increases area by a near negligible amount. 

\subsection{Reliability}

We use LTSPICE simulations to confirm that our shift operation works reliably. To study the effects of process variation on shifting, our simulation models variation in the cell capacitance, transistor length/width, and the bitline/wordline capacitance and resistance. We implement the 22nm low-power transistor models \cite{websitekey} and use cell/transistor parameters pulled from our VLSI layout for a 22nm technology node (cell capacitance = 25fF; transistor length/width = 22nm/44nm). 

We use Monte-Carlo simulations to understand the impact process variation has on shifting. We increase the process variation from 0 to $\pm$20\% and run 100,000 simulations for each level of process variation. Table 4 shows the percentage of iterations in which our shifting operation fails for each level of variation. 

\begin{table}
\centering
\footnotesize
\begin{tabular}{| l | l | l | l | l |}
\hline
\textbf{Variation} & ±0\% & ±5\% & ±10\% & ±20\% \\
\hline
\textbf{\%Failures} & 0.00\% & 0.5\% & 14\% & 30\% \\
\hline

\end{tabular}
\caption{Effect of Process Variation on Shift}
\end{table}

\subsection{Area Overhead}
In this section, we analyze the area overhead of our migration cell-based shifting approach and compare it with existing PIM architectures that support bit-shifting operations.

\subsubsection{Migration Cell Design Overhead}
Our migration cell-based design introduces minimal area overhead to the DRAM subarray. As discussed in section 3, a migration cell can be made between two cells simply by connecting the nodes of the top plates of each storage capacitor with a wire. We do not change the existing structure in any other way, and there is no additional control wiring or logic needed. This extra wiring introduces <1\% area overhead based on estimates from Lu et al. \cite{lu2015improving} and our VLSI layout. This migration cell design can be implemented on top of Ambit \cite{seshadri2017ambit} which would incur an additional estimated 1\% area overhead. 

\begin{table}
\centering
\footnotesize
\begin{tabular}{| l | l | l |}
\hline
\textbf{Design} & \textbf{Added Circuitry} & \textbf{Area Overhead} \\
\hline
w/ Migration Cells & Wiring & <1\% (without Ambit) \\
\hline
SIMDRAM & Control unit + Transposition unit & 0.2\% (vs Intel Xeon CPU) \\
\hline
DRISA 3T1C & Shifters, controllers, bus, buffers & $\sim$6.8\% (vs 8Gb DRAM) \\
\hline
DRISA 1T1C-nor & NOR gates + latches + shifters & $\sim$34\% added circuits \\
\hline
DRISA 1T1C-mixed & Mixed logic gates + shifters & $\sim$40\% added circuits \\
\hline
DRISA 1T1C-adder & Adders + shifters & $\sim$60\% added circuits \\
\hline

\end{tabular}
\caption{Area Overhead of PIM Architectures}
\end{table}

\subsubsection{Comparison with Prior Work}
Table 5 shows the area overhead of our design compared with existing PIM architectures that are capable of bit-shifting operations. SIMDRAM \cite{hajinazar2021simdram} uses a vertical data layout combined with a transposition unit in the memory controller rather than adding logic to the DRAM die in order to achieve minimal DRAM chip overhead. However, the data transposition between horizontal and vertical layouts results in significant runtime overhead, makes system integration complicated, and requires dedicated transposition hardware.  Our design operates directly on horizontally-stored data, thus eliminating these transposition costs and maintaining comparable area overhead ($\sim$1-2\% vs $\sim$1\%). DRISA \cite{li2017drisa} enables horizontal data movement by adding dedicated shifter circuits beneath the sense amplifiers. DRISA offers a 3T1C design and several 1T1C variants. The 1T1C variants add logic gates, latches, and even full adders implemented in DRAM process technology. These 1T1C variants result in 34-60\% area overhead. Their 3T1C design uses 3T1C cells for cell based computing, but this incurs a 6.8\% overhead due to the larger cell size (30F\(^2\)  compared to standard 6F\(^2\)). The DRISA paper concluded that building complex logic circuits with DRAM process technology is inefficient and not feasible.    

\subsubsection{Advantages of Our Approach}
Our migration cell-based design achieves favorable area overhead through several advantages: 
\begin{itemize}
    \item Minimal additional circuitry: Unlike DRISA which adds complex logic circuits, we simply add wiring and leverage existing cell structures to create the migration cells.  
    \item No data transposition overhead: Unlike SIMDRAM which adds a transposition unit and incurs energy, runtime, and complexity costs, our design operates on standard horizontal data layouts. 
    \item Leveraging existing structures: We eliminate the need for new circuitry and logic by using existing cell structures with additional wiring, reducing area overhead impact and design complexity.  
\end{itemize}

Our design incurs $\sim$1-2\% area overhead which positions it between SIMDRAM's minimal DRAM overhead (at the cost of transposition) and DRISA's higher performance (at the cost of heavy area overhead). For applications requiring frequent shift operations—such as shift-and-add multiplication \cite{de2024count2multiply}, carry propagation in adders, and Galois field arithmetic for AES encryption \cite{reed1960polynomial} and Reed-Solomon error correction \cite{wicker1999reed}—the minimal area overhead of our design allows for practical deployment without the inefficiencies of complex logic circuits and penalties of data transposition.  

\begin{figure}
    \centering
    \includegraphics[width=0.75\linewidth]{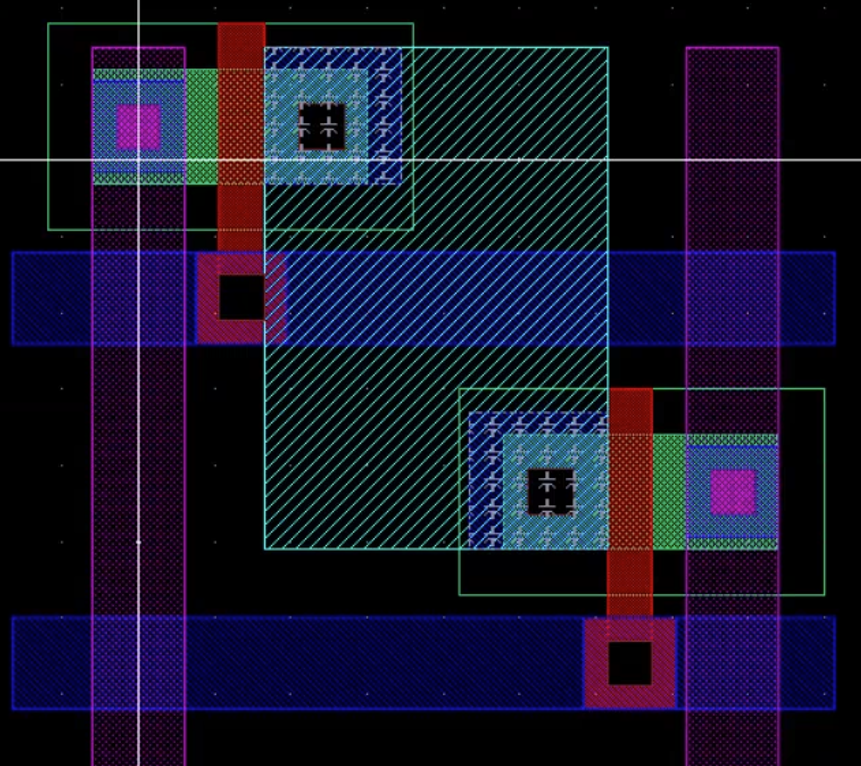}
    \caption{VLSI layout of a Migration Cell}
    \label{fig:placeholder}
\end{figure}

\section{VLSI Layout}

We used Cadence Virtuoso to implement the physical layout of our migration cell design in a 22nm technology node. The migration cell consists of two standard cells of the 1T1C structure. Each cell layout measures the minimum width and length for 22nm technology (width = 0.044\unit{\us} and length = 0.022\unit{\us}). We made sure to maintain compatability with standard DRAM cell pitch constraints for wordlines and bitlines at 22nm technology \cite{auth201222nm}. We use metal 1 for the wordlines and metal 2 for the bitlines. Figure 4 shows our migration cell layout. 

We use Metal-Insulator-Metal (MIM) capacitors. A storage capacitor at 22nm typically has a capacitance of $\sim$25fF. In order to obtain this desired capacitance, our plates of the MIM capacitor must have a certain area, A, and must be split apart by a certain distance, d, for the dielectric in the middle to be inserted. We choose hafnium oxide ($HfO_2$) as the dielectric based on prior work with 22nm technology \cite{auth201222nm}. $HfO_2$ has a dielectric constant $\varepsilon$$_r$= 20 \cite{park2006enhancement}. The dielectric thickness of $HfO_2$ is estimated to be in the range of 6-10nm \cite{mondon2003electrical}. Based on these values and the permittivity of free space ($\varepsilon$$_0$ = 8.8854 × $10^{-12}$F/m), we calculate the area of each plate of our MIM capacitor to be 1,129,000 $nm^2$ (or 1.129 × $10^6$ $nm^2$). We take the square root of this value to get the side length = 1,063 $nm$ (or 1.06\unit{\us}). Each MIM capacitor plate in our layout would have square dimensions: 1.06\unit{\us} × 1.06\unit{\us}. 
\section{Results and Discussion}
\label{sec:Results and Discussion}
Our evaluation demonstrates that the migration cell-based in-DRAM shifting design provides an efficient approach to bit-shifting in open-bitline architectures. The results validate our design through energy efficiency, performance, reliability, and physical implementation. 

Our simulations using NVMain show that a shift operation achieves energy efficiency of $\sim$4 nJ/KB, and the energy per shift stays constant at 31-32 nJ. We observed that the latency per shift operation ($\sim$208 ns) aligns with the baseline cost of 4 AAP commands. This time is also consistent with other charge-sharing-based PIM operations. Our approach eliminates data transfers between DRAM and the CPU completely resulting in a 40-60× reduction in energy consumption. 

SIMDRAM achieves fast single-operation shifts but incurs substantial transposition overhead (1000-10,000 nJ for large operands), but our design operates on horizontally-stored data resulting in no transposition overhead. DRISA adds dedicated shifter circuits at various costs (ranges from 6.8-60\%) of area overhead depending on which variant is used, but our design achieves comparable energy efficiency (4 nJ/KB vs 5-20 nJ/KB) and maintains minimal area overhead of <1-2\% by leveraging the existing cell structures and adding wiring to make the migration cells. 

Our LTSPICE simulations validate the circuit-level correctness and reliability of the shift operation under realistic conditions at a 22nm technology node. The Monte Carlo analysis showed that our design operates reliably under nominal conditions (0\% failure rate at $\pm$0\% variation). It also maintains acceptable reliability under moderate process variation (0.5\% failure rate at $\pm$5\% variation, and 14\% failure rate at $\pm$10\% variation). This negative effect of process variation at a smaller technology node like 22nm is expected \cite{kang2014co}. The higher variation levels show increased failure rates, but these extreme variations are uncommon in modern DRAM manufacturing \cite{chang2016understanding}. 

The Cadence Virtuoso VLSI layout at 22nm technology shows that our design is physically realizable. The migration cells require only addition routing between existing cells in the standard subarray structure. Our layout follows all foundry design rules such as minimum width/spacing and pitch constraints \cite{auth201222nm} to ensure it can be manufacturable. 

Our design offers a combination of advantages to existing DRAM architectures such as minimal area overhead, compatibility with standard DRAM structures and operations, and energy-efficient operation. Our design can be implemented on top of Ambit which results in a functionally complete set of bulk bitwise operations and shifting added which is important for supporting complex PIM applications that require logical operations and bit-shifting such as arithmetic operations (adders and multipliers) and cryptographic algorithms (AES).  

\section{Future Work}
\label{sec:Related Work}
While our current work establishes an efficient design for in-DRAM bit-shifting, there are several promising options for future exploration. 

\subsubsection{Adder Circuit Analysis} - A natural extension of this work is to analyze any performance improvements our shifting operation enables for arithmetic circuits such as Ripple carry adders and Kogge-Stone adders. We would quantify the speedup and energy savings achievable by using shifting where necessary when implementing these adders in memory. 

\subsubsection{Cryptographic Applications} - Galois field arithmetic forms the mathematical foundation for AES encryption and Reed-Solomon error correction codes \cite{wicker1999reed} \cite{reed1960polynomial} and relies heavily on bit-shifting for the polynomial multiplication and reduction steps. A detailed case study implementing these cryptographic algorithms entirely in memory using PIM operations and shifting would demonstrate real-world applicability. 

\subsubsection{Multi-Bit Shift Extensions} - Our current design allows only single bit shifts at a time. More bit shifts require multiple shift operations. Future work could explore extensions to this design that enables multi-bit shifts in fewer operations. This would benefit applications requiring large shift distances. 

\section{Conclusion}
\label{sec:Conclusion}

PIM architectures enable computation directly within memory to help combat the memory wall problem. Bit-shifting has presented itself to be challenging to implement efficiently in memory due to the isolation of adjacent bitlines. Prior approaches to shifting required substantial are overhead for dedicated shifter circuits (DRISA) or substantial transposition overhead using vertical data layouts (SIMDRAM). 

In this paper, we proposed a novel DRAM subarray design that enables in-DRAM bit-shifting for open-bitline architectures. Our design enables horizontal data movement by adding migration cells in the top and bottom rows of each subarray. A shift operation is equivalent to a sequence of 4 AAP commands. Our design remains compatible with standard DRAM operations and eliminates the need for complex logic by leveraging existing cell structures to make the migration cells with minimal wiring added. 

Our evaluation shows the efficiency and reliability of our design. Our shift operation achieves energy efficients of $\sim$4 nJ/KB and a latency of $\sim$208 ns per shift. This shows a 40-60× energy reduction when compared to conventional movement between CPU and DRAM. Circuit-level LTSPICE simulations validate reliably operation across various technology nodes and process variations. Our Cadence Virtuoso VLSI layout at 22nm technology node demonstrates physical feasibility and realizability. The design follows all design rules and incurres <1-2\% area overhead which is significantly lower than prior approaches like DRISA (6.8-60\%)

By enabling efficient bit-shifting on horizontally stored data, our design opens new possibilities for accelerating PIM applications such as Ripple carry adders, Kogge-Stone adders, and Galois field arithmetic used in cryptographic algorithms like AES. Our design has minimal area overhead and is compatible with existing PIM frameworks like Ambit which makes it practical for use on future memory systems.  

\bibliographystyle{ACM-Reference-Format}
\bibliography{sample-base}


\end{document}